\documentclass[12pt]{revtex4}

\usepackage{graphicx,epsfig}
\usepackage{bm,longtable,enumerate}
\usepackage{amsmath,amssymb}

\def\beq{\begin{equation}}
\def\eeq{\end{equation}}
\def\beqa{\begin{eqnarray}}
\def\eeqa{\end{eqnarray}}

\begin{document}

\title{D-dimensional  Randall-Sundrum models from Brans-Dicke theory and Kaluza-Klein modes}

\author{C.~A.~Ballon Bayona}
\email{ballon@if.ufrj.br}
\affiliation{
Instituto de F\'{i}sica, Universidade Federal do Rio de Janeiro,\\
Caixa Postal 68528, RJ 21941-972, Brazil}

\author{
Cristine N. Ferreira}

\email{crisnfer@pq.cnpq.br}

\affiliation{N\'ucleo de Estudos em F\'{\i}sica, Instituto Federal de  Educa\c{c}\~ao, \\ Ci\^encia e Tecnol\'ogia Fluminense
28030-130 Campos, RJ, Brazil }

\begin{abstract}
We investigate the spectroscopy of scalar and vector Kaluza-Klein modes that arise in a deformed Randall-Sundrum model that is constructed from Brans-Dicke theory. The non-minimal coupling in the Brans Dicke theory translates into a deformation of the Randall-Sundrum geometry that depends on the Brans-Dicke parameter $\omega$. We find that $ \omega$ parameter has  a non-trivial effect in the spectroscopy of scalar and vector Kaluza-Klein modes. Our results suggest the interpretation of $\omega $ as a fine-tuning parameter. 

\end{abstract}


\maketitle

\section{Introduction}
The hierarchy between gravitational and electromagnetic forces motivated in the early times the Dirac cosmological model \cite{Dirac:1937ti} that considers a time dependent gravitational constant. This model inspired some field theory approaches like the Jordan Model \cite{Jordan:1949ti} in which the gravitational constant is taken as a function of some scalar field. A complete scalar-tensor theory of gravitation was proposed in 1961 by Brans and Dicke where the gravitational constant is inversely related to the scalar field \cite{BransDicke}.

Kaluza-Klein theories and String theory motivated several models involving extra dimensions and branes being the most interesting the one proposed by  Randall and Sundrum \cite{Randall:1999ee}. This model  considers a configuration of two  4D branes in a 5D space-time  with negative cosmological constant. The hierarchy problem between the Planck and electro-weak scale is solved by the warp factor present in the 5d metric. An important problem in the Randall-Sundrum scenario is the fixing of the extra dimension size $L$. The first attempt to fix $L$ was to consider a five dimensional scalar field with brane potentials \cite{Goldberger:1999uk}. Including the backreaction of this field on the metric led to a five dimensional scalar-tensor model \cite{DeWolfe:1999cp} that differs from the original Randall-Sundrum solution. Recently, a five dimensional Brans-Dicke model with branes was proposed in \cite{Mikhailov:2006vx}. Working in the Jordan-Fierz frame, the 5D Brans-Dicke action can lead to  metric solutions very similar to the original Randall-Sundrum metric. The  model of \cite{Mikhailov:2006vx} includes backreaction and the solution is stable because the size of the extra dimension is fixed by the scalar field. 

In this paper we construct D-dimensional Randall-Sundrum models from Brans-Dicke theory. We consider a BPS-like mechanism that translates the second-order differential equations coming from the Brans-Dicke action into first-order ones. This way we find a special class of scalar potentials that simplifies the background solutions. A particular choice of the scalar potential leads to a  Randall-Sundrum solution for the metric which can be stabilized following a procedure similar to \cite{Mikhailov:2006vx}. We analyze the   possible implications of the D dimensional Brans-Dicke parameter  by performing a Kaluza-Klein decomposition of a massless scalar fluctuation living in the bulk. We find an interesting dependence of the $D-1$ dimensional scalar masses on the D dimensional Brans-Dicke parameter. We also discuss the effect of the Brans-Dicke parameter on the Kaluza-Klein modes arising on a recent Higgless model for electroweak symmetry breaking \cite{Csaki:2003zu}. Our results suggest the possibility of considering the Brans-Dicke parameter as a fine-tuning for the $W$ and $Z$ resonances. 

We begin in Sec. II with a review of the Randall-Sundrum metric. In Sec. III we show how this metric arises from the Brans-Dicke theory via a BPS-like mechanism. In Sec. IV we analyze the Kaluza-Klein modes coming from the decomposition of a massless scalar fluctuation while in Sec. V we discuss the gauge field Kaluza-Klein modes of a Higgless electroweak model. We end with
conclusions in Sec. VI.

\section{The Randall-Sundrum Metric in D-dimensions}
The Anti-de-Sitter space-time is a maximally symmetric solution of the Einstein equations with negative cosmological constant $\Lambda$. This space
-time can be interpreted as a hyperboloid of radius $\ell$ related to the cosmological constant by $ -\Lambda \, \ell^2 = (D-1)(D-2)$. The Poincar\'e chart cuts the hyperboloid in two regions (see \cite{Bayona:2005nq} for details). The metric of each region can be written as   
\begin{equation}
d \tilde s^2 = {1 \over k^2 z^2}[-dt^2 + d\bar x^2 + dz^2]\, ,
\end{equation}
where $k=1/\ell$, \, $ d\bar x^2 = \sum_{i=1}^{D-2} dx_i^2$ and $z>0$ (or $z<0$). The Randall-Sundrum metric can be constructed by considering two slices of the $z>0$ region. For this purpose, it is convenient to define a new coordinate $\Omega$ by $z =  {1 \over k}e^{k | \Omega |}$. The two AdS slices are given by $0<\Omega\le L$ and $-L\le \Omega<0$ and can be joined at $\Omega=0$.  The relation between $z$ and $\Omega$ is plotted in Fig.1. The metric in terms of $\Omega$ reads 

\begin{equation}
d\tilde s^2 = e^{2 \sigma(\Omega)}[-dt^2 + d\bar x^2] + d\Omega^2 \, ,\label{RandSundmetric}
\end{equation}
\noindent where $\sigma(\Omega)=-k|\Omega|$ and $-L\le\Omega\le L$. Identifying $\Omega$ with $-\Omega$ we get the orbifold space $S^1/Z_2$. The metric (\ref{RandSundmetric}) naturally satisfies this condition.

\begin{figure}
\centering
\includegraphics[width=7.0cm]{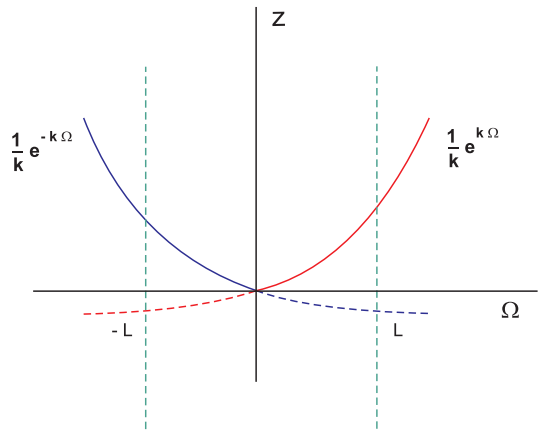}
{\caption{The z dependence on  $\Omega $ . 
 }\label{Fig1}}
\end{figure}

The Randall-Sundrum metric was obtained from Einstein equations coming from a D-dimensional gravitational action with negative cosmological constant in the presence of two (D-1)-branes located at $\Omega=0$ and $\Omega=L$ with opposite tensions. We will see in the next section how this metric also arises from a D-dimensional Brans-Dicke theory.

\section{Brans-Dicke theory and the deformation of the Randall-Sundrum geometry}

In this section we will use a BPS-like mechanism to solve the field equations of motion coming from a D-dimensional Brans-Dicke theory with two (D-1)-brane potentials. In this theory there is a scalar field non-minimally coupled to gravity. The total action is given by   
\begin{eqnarray}
S &=& \int d^{D-1}x d\Omega  \sqrt{-\tilde g}\Big[\tilde \Phi {\tilde R} - {\omega \over \tilde \Phi} {\tilde g}^{MN} \partial_M\tilde \Phi \partial_N \tilde\Phi - \tilde V(\tilde \Phi) \Big] \cr
& -& \int_{_{\Omega=0}} \!\! \!\! d^{D-1}x \sqrt{- \tilde h}\tilde \lambda_1(\tilde \Phi)  - \int_{_{\Omega=L}} \!\! \!\! d^{D-1}x \sqrt{- \tilde h} \tilde \lambda_2(\tilde \Phi)  \, , \label{totalaction}
\end{eqnarray}
\noindent where the coordinates $ x^M=(x^{\mu}, \Omega) $ consist on D-1 non-compact coordinates  $x^{\mu}$  and a compact coordinate $\Omega$ defined in the interval $-L\le\Omega\le L$ with the identification $\Omega \to -\Omega$. The Ricci scalar of the metric ${\tilde g}_{MN}$ is denoted by ${\tilde R}$ and we work with the signature (-, +,..,+).We denote by $\tilde h_{\mu \nu} $  the induced metric on the branes. The term $\tilde V(\tilde \Phi)$ is a bulk potential while $ \tilde \lambda_1$, $ \tilde \lambda_2 $ are brane potentials.  The constant $ \omega$ is the D-dimensional Brans-Dicke parameter. The orbifold condition in $\Omega$ implies 
\begin{eqnarray}
{\tilde g}_{\mu \nu}(x, - \Omega) &= & {\tilde g}_{\mu \nu}(x, \Omega)\qquad  \quad ; \qquad {\tilde g}_{\Omega \Omega}(x, - \Omega)  =  {\tilde g}_{\Omega \Omega}(x, \Omega) \, ,\nonumber\\
{\tilde g}_{\mu \Omega}(x, - \Omega)&= & -{\tilde g}_{\mu\Omega}(x, \Omega) \qquad  ; \qquad {\tilde \Phi}(x, - \Omega)  = {\tilde \Phi} (x, \Omega) \, .
\end{eqnarray}

The action (\ref{totalaction}) leads to the following background equations 
\begin{eqnarray}
\tilde R^{\Omega \Omega} - {1 \over 2} \tilde g^{\Omega \Omega}\Big( \tilde R - {\tilde V \over \tilde \Phi }\Big) 
+ \, \, {\omega \over \tilde \Phi^2} \partial_{M}\tilde \Phi \partial_N \tilde \Phi\Big[ {1 \over 2} \tilde g^{\Omega \Omega}\tilde g^{M N} - \tilde g^{M \Omega}\tilde g^{N \Omega} \Big] \cr 
+ \, \, {\tilde \Phi_{;M;N} \over \tilde \Phi}\Big[\tilde g^{\Omega  \Omega}\tilde g^{ MN } - \tilde g^{M \Omega}\tilde g^{N \Omega}\Big] \, = \, 0 \, ; \\
\tilde R^{\mu \nu} - {1 \over 2} \tilde g^{\mu \nu}\Big( \tilde R - {\tilde V \over \tilde \Phi }\Big) 
+ \, \, {\omega \over \tilde \Phi^2} \partial_{M}\tilde \Phi \partial_N \tilde \Phi\Big[ {1 \over 2} \tilde g^{\mu \nu}\tilde g^{M N} - \tilde g^{M \mu}\tilde g^{N \nu} \Big] \cr 
+ \, \, {\tilde \Phi_{;M;N} \over \tilde \Phi}\Big[\tilde g^{\mu  \nu}\tilde g^{ MN } - \tilde g^{M \mu}\tilde g^{N \nu}\Big] 
+ \, \, {1 \over 2} {\tilde g^{\mu \nu}\over \sqrt{g_{\Omega \Omega}} \tilde \Phi}\tilde \lambda_2 \delta(\Omega -L) + {1 \over 2} {\tilde g^{\mu \nu}\over \sqrt{g_{\Omega \Omega}} \tilde \Phi}\tilde \lambda_1 \delta(\Omega )  = \, 0 \, ; \\
{\omega \over \tilde \Phi^2} \partial_M \tilde \Phi \partial_N \tilde \Phi \tilde g^{MN} \! \!+ \! \!{2 \over \sqrt{-\tilde g}}\partial_M\Big[\sqrt{-\tilde g}{\omega \over \tilde \Phi}\tilde g^{MN} \partial_N \tilde \Phi\Big] \! \!+ \!  \tilde R   - {\partial \tilde V \over \partial \tilde \Phi} \cr 
- {1 \over \sqrt{\tilde g_{\Omega \Omega}}}{ \partial \tilde \lambda_2 \over \partial \tilde \Phi} \delta(\Omega -L) - {1 \over \sqrt{\tilde g_{\Omega \Omega}}}{ \partial \tilde \lambda_1 \over \partial \tilde \Phi} \delta(\Omega )=0 \,.
\end{eqnarray}

We consider  the following ansatz for the metric and scalar field : 
\begin{equation}
ds^2 = e^{2\sigma(\Omega)}\eta_{\mu \nu} dX^{\mu}dX^{\nu} + d \Omega^2 \quad , \quad \tilde \Phi = \tilde \Phi(\Omega) \, ,\label{ansatzBD}
\end{equation}
where $\tilde \Phi(\Omega)$ and $\sigma(\Omega)$ are even functions in $\Omega$. The background equations above then translates into  a system of second order differential equations 

\begin{eqnarray}
&&{1 \over 2}(D-2)(D-1) \sigma'^2 \tilde \Phi + {\tilde V \over 2}  -  {w \over 2\tilde \Phi} \tilde \Phi'^2 + (D-1) \sigma' {\tilde \Phi}' = 0 \, ;  
\label{einsteinV1} \\
&&\tilde \Phi'' + {w \over \tilde \Phi} \tilde \Phi'^2 - \sigma' \tilde \Phi' + (D-2) \sigma'' \tilde \Phi 
+ {1 \over 2} \tilde \lambda_1 \delta(\Omega) + {1 \over 2} \tilde \lambda_2 \delta(\Omega -L) =0\, ; \label{einsteinV2} \\
&&{w \over \tilde \Phi}\tilde \Phi'' + (D-1)w{\sigma'\over \tilde \Phi} \tilde \Phi' -(D-1)\sigma'' 
 -{1\over2}D(D-1)\sigma'^2-{w \over 2\tilde \Phi^2}  \tilde \Phi'^2 \cr 
&&-{1 \over 2} {\partial \tilde V \over \partial \tilde \Phi}  
- {1 \over 2} {\partial \tilde \lambda_1 \over \partial \tilde \Phi}\delta(\Omega)- {1 \over 2} {\partial \tilde \lambda_2 \over \partial \tilde \Phi}\delta(\Omega -L)=0  \label{einsteinV3} \,.
\end{eqnarray} 

Finding a solution of these differential equations is in general complicated for an arbitrary potential $\tilde V (\tilde \Phi)$. We could also invert the problem and solve the equations for the  scalar field solution and potential once we know the metric . In this work we use a BPS-like mechanism that simplifies the background equations and leads to a special class of potentials. The Randall-Sundrum solution for the metric arises from a particular potential belonging to this class.    

If we substitute the ansatz (\ref{ansatzBD}) in the lagrangian density of eq. (\ref{totalaction}) we find 
\begin{equation}
{\cal L} = -e^{(D-1) \sigma}\{(D-1)\tilde \Phi (2\sigma ''  + D\sigma '^2)  + \omega {\tilde \Phi'^2 \over \tilde \Phi}  + \tilde V + \tilde \lambda_1 \delta (\Omega) + \tilde \lambda_2 \delta (\Omega-L) \}.
\end{equation}

In order to have periodicity in the coordinate $\Omega$ and justify the presence of the $\delta(\Omega-L)$ function the lagrangian density has to be integrated from $-L+\epsilon$ to $L+\epsilon$ and make $\epsilon \to 0$ at the end.  The lagrangian density can be rewritten in the following form 
\begin{eqnarray}
{\cal L} &=&  -e^{(D-1)\sigma} \Big \{ (\omega + {D-1 \over D-2})\tilde \Phi\Big({\tilde \Phi' \over \tilde \Phi} - \hat{W} + (D-2) \tilde \Phi  {\partial \hat{W} \over \partial \tilde \Phi } \Big)^2  \cr
&-& (D-1) (D-2)\tilde \Phi  \Big( \sigma' + {1 \over D-2} {\tilde \Phi' \over \tilde \Phi} -(\omega + {D-1 \over D-2})  \hat{W} \Big)^2  \cr
&+& \Big[ \tilde V +  [(D-2)\omega +(D-1)][((D-1)\omega+D)\tilde \Phi \hat{W}^2 
+   \tilde \Phi^2 \hat{W} {\partial \hat{W} \over \partial \tilde \Phi} - (D- 2)  \tilde \Phi^3\left({\partial \hat{W} \over \partial \tilde \Phi}\right)^2]\Big] \cr 
&+& \Big[ 2[(D-2)\omega +  (D-1)] {\partial \hat{W} \over \partial \Omega} \tilde \Phi + \tilde \lambda_1 \delta(\Omega) +\tilde \lambda_2\delta(\Omega-L) \Big]\Big \} \cr 
&-& 2(D-1)\Big[\sigma'\tilde \Phi e^{^{(D-1)\sigma}} \Big]' + 2[(D-2)\omega + (D-1)]\Big[\tilde \Phi \hat{W} e^{^{(D -1)\sigma}}\Big]' , \label{lagrangiandensity}
\end{eqnarray}
where we have introduced an arbitrary odd function 
\begin{equation}
\hat{W}(\tilde \Phi)=\left\{\begin{array}{rc}
W(\tilde \Phi) &\mbox{if}\quad 0< \Omega< L \, ,\\
-W(\tilde \Phi) &\mbox{if}\quad -L< \Omega < 0\, .
\end{array}\right.  \label{wchapeu}
\end{equation}

The last two terms in (\ref{lagrangiandensity}) are total derivatives so they vanish using the periodicity of $\Omega$.  The first two terms are square terms which are zero when  
\begin{eqnarray}
 \tilde \Phi' &=& [ \, \tilde \Phi \hat{W} -(D-2) \, \tilde \Phi^2 {\partial \hat{W} \over \partial \tilde \Phi}] \, ;\label{firstorder1} \\ \sigma' & =& [(\omega +1)\, \hat{W}  + \,\tilde \Phi {\partial \hat{W} \over \partial \tilde \Phi }] \, .\label{firstorder2}
\end{eqnarray}

Assuming that the equations above are satisfied by the scalar field and the metric we find that the following class of bulk potentials 
\begin{eqnarray}
\tilde V &=&- [(D-2)\omega +(D-1)][((D-1)\omega+D)\tilde \Phi \hat{W}^2 \nonumber \\ 
& & +2\tilde \Phi^2\hat{W} {\partial \hat{W} \over \partial \tilde \Phi} - (D-2) \tilde \Phi^3({\partial \hat{W} \over \partial \tilde \Phi})^2] \, , \label{classpotential}
\end{eqnarray}
\noindent with the brane conditions 
\begin{equation}
 \! 2[(D-2)\omega + (D-1)] {\partial \hat{W} \over \partial \Omega}  
=-{\tilde \lambda_1 \over \tilde \Phi} \delta(\Omega) - {\tilde \lambda_2 \over \tilde \Phi} \delta(\Omega-L)\, ,
\end{equation}
\noindent lead to a vanishing action. Using (\ref{wchapeu}) the brane conditions read 
\begin{eqnarray}
&&[(D-2)\omega + (D-1)] W( \tilde \Phi) |_{\Omega=0^+} = -{\tilde \lambda_1 \over 4 \tilde \Phi}\, ; \label{branecond1} \\
&&[(D-2)\omega + (D-1)] W( \tilde \Phi) |_{\Omega=L^-} = {\tilde \lambda_2 \over 4 \tilde \Phi}\, , \label{branecond2}
\end{eqnarray}
\noindent and similar for the derivatives in $\tilde \Phi$. It is straightforward to show that the system of equations (\ref{firstorder1})-(\ref{branecond2}) give background solutions that also satisfy the background equations (\ref{einsteinV1})-(\ref{einsteinV3}). This way we find a BPS-like mechanism that gives background solutions for second order differential equations  by solving first order equations that appear inside the square terms in the lagrangian density. Because the square terms appear with opposite signs there is no a Bogomolnyi bound. This mechanism is similar to that found in ref. \cite{DeWolfe:1999cp}. Note that for the case $D=5$ our equations (\ref{firstorder1})-(\ref{branecond2}) reduce to those obtained in \cite{Mikhailov:2006vx}.

A Randall-Sundrum solution for the metric is obtained for the case of constant $W$ where the potential and background solutions reduce to 
\begin{eqnarray}
\tilde V(\tilde \Phi ) = \Lambda \tilde\Phi \quad ; \quad \sigma = - k |\Omega| ;\label{sigma}\\ 
\tilde \Phi = C \exp({\sigma \over \omega +1})  ,\label{RSbackgroundsolution}
\end{eqnarray}
with 
\begin{eqnarray}
C &=& {1 \over 16 \pi G_D} \quad  ;  \quad W = - {k \over (w +1)} \, \cr 
\Lambda &=& - [(D-2)\omega +(D-1)][(D-1)\omega+D)] W^2 \label{RSbackgroundsolution2}
\end{eqnarray}

\noindent The value of $C$ was chosen for convenience. The brane potentials in this case are  
\begin{equation}
\tilde \lambda_1 =  \lambda \tilde \Phi \quad ; \quad \tilde \lambda_2  =  -\lambda \tilde \Phi\, ,
\end{equation}

\noindent with $ \lambda =4 \sqrt{{(D-2)w +(D-1) \over (D-1) w +D}}\sqrt{- \Lambda}$.

Note that although we have obtained the Randall-Sundrum metric (\ref{RandSundmetric}),
the scenario given by eqs.(\ref{sigma})-(\ref{RSbackgroundsolution2}) is different from the 
traditional Randall-Sundrum scenario because the metric couples non-minimally with a non-trivial 
background scalar field. The traditional Randall-Sundrum scenario can be obtained in the limit $\omega \to \infty$
in which the scalar field becomes trivial, as discussed in \cite{Mikhailov:2006vx}. 

\medskip

{\bf The Einstein frame}

\medskip
If we perform the following background transformations :

\begin{eqnarray}
\tilde g_{M N}& = & e^{2 \alpha \Phi} g_{MN}\quad ; \quad \tilde \Phi  =  {1 \over 16\pi G_D} e^{-(D-2) \alpha \Phi} \, ;\cr
\tilde V(\tilde \Phi) &=& e^{-D\alpha \Phi} V(\Phi)\quad ; \quad \tilde \lambda_i(\tilde \Phi) =  e^{-(D-1)\alpha \Phi} \lambda_i(\Phi) \, ,
\end{eqnarray}
\noindent with 
\begin{equation}
\alpha^2  =  {1\over 32\pi G_D(D-2)^2}\Big[w + {D-1 \over  D-2}\Big]^{-1} \, ,
\end{equation}

\noindent we go from the Jordan-Fierz frame (in which the Brans-Dicke theory is originally formulated) to the Einstein frame. These background transformations are known in the literature as conformal transformations \cite{Faraoni}.  Note that this transformation imposes a reality condition for the Brans-Dicke parameter: $w >  - {D-1 \over  D-2} $. 
The total action (\ref{totalaction}) becomes
\begin{eqnarray}
S &=& \int\! d^Dx \sqrt{-g}\Big({1 \over 16 \pi G_D}R -{1 \over 2} g^{MN}\partial_M  \Phi \partial_N  \Phi - V(\Phi )\Big) \cr
&-&  \int_{\Omega =0} \! \! \!\! \! \! d^{D-1}x \sqrt{-h} \lambda_1(\Phi)  -\int_{\Omega=L}\!\! \!\! d^{D-1}x \sqrt{-h}  \lambda_2(\Phi) \, .
\end{eqnarray}

In the Einstein frame, the background solutions of (\ref{RSbackgroundsolution}) become  
\beqa
ds^2 &=&  e^{{2 \sigma \over (D-2)( w +1)}}\! \Big[  e^{2 \sigma}\eta_{\mu \nu}dx^{\mu }dx^{\nu} \! +  \!d \Omega^2\! \Big]\, ; \cr
\Phi &=& -{1 \over  (D-2)\alpha}\Big[ {\sigma \over w +1}\Big]\, . \label{backgroundeinstein}
\eeqa
In terms of the coordinate $z = {1 \over k}e^{-\sigma(\Omega)}$ the metric reads
\begin{equation}
ds^2 = f_\omega(z) \, \frac{1}{(k z)^{2}} [\eta_{\mu \nu} dx^{\mu} dx^{\nu} + dz^2] \quad ;\quad f_\omega(z) \equiv \left( k z \right)^{{-2\over (D-2)(w+1)}}\, . \label{zmetric}
\end{equation}
This metric can be interpreted as a deformed Randall-Sundrum metric where the deformation is given by $f_\omega(z)$. Note that the Planck brane is localized at $z=1/k$ while the TeV brane is localized at $z=(1/k)e^{kL}$.  In the limit $\omega \to \infty$ the deformation factor $f_\omega(z)$ goes to $1$ and we recover the original Randall-Sundrum metric. 

\section{Spectroscopy of  scalar Kaluza-Klein Modes }

Now we consider the compactification of a massless scalar field fluctuation in the Einstein frame. This frame is well motivated for many reasons being the most important the positive sign of the energy density \cite{Faraoni}. A scalar field fluctuation can be described by the following action
\begin{equation}
S = -{1 \over 2} \int d^{^{D-1}}\!\!\!x \int d \Omega  \sqrt{-g} g^{M N} \partial_{M} \varphi \partial_{N} \varphi \, .\end{equation}
\noindent This action can be decomposed as 
\begin{equation}
S = -{1 \over 2}\int d^{^{D-1}}\!\!\!x \int d \Omega \Big[\sqrt{-g} h(\Omega) \eta^{\mu \nu} \partial_{\mu} \varphi \partial_{\nu} \varphi - \varphi \partial_{\Omega}\Big( \sqrt{-g}g^{\Omega \Omega}\partial_{\Omega} \varphi\Big)\Big] \, ,
\end{equation}

\noindent where $h(\Omega)$ is defined by  $ g^{\mu \nu}=\eta ^{\mu \nu}h(\Omega)$. The Kaluza-Klein decomposition of $\varphi(x, \Omega)$ is 
\begin{equation}
\varphi(x, \Omega) = {1 \over \sqrt{L}}\sum_n \phi_n(x) \chi_n(\Omega)\label{expan1} \, .
\end{equation}

If the modes $\chi_n(\Omega)$ satisfy the relations 
\begin{eqnarray}
{1 \over L} \int_{-L}^{L} d \Omega  \sqrt{-g} h(\Omega)\chi_n(\Omega) \chi_m(\Omega)=  \delta_{nm}\, ; \label{normalizationcond}\\
\frac{d}{d \Omega} \Big( \sqrt{-g}g^{\Omega \Omega} \frac{d \chi_n}{d\Omega}\Big) = - m_n^2 \sqrt{-g} h(\Omega) \chi_n  \, ,\label{eqbess}
\end{eqnarray}

\noindent then we get the D-1 dimensional action for $\phi_n(x)$ :

\begin{equation}
S_{eff} = -{1\over 2}\sum_n\int d^{^{D-1}}\! \! \!  x \, \Big[\eta^{\mu \nu} \partial_{\mu} \phi_n \partial_{\nu} \phi_n+ m^2_n \phi_n^2 \Big]\, .
\label{dialton4d} 
\end{equation}
As in usual Kaluza-Klein compactifications, the bulk field $\phi(x, \Omega) $ manifests to a $D-1 $ dimensional observer as an infinite  "{\it tower}" of scalars $\phi_n(x)$ with masses $m_n$.  

The tower of masses $m_n$ can be obtained by solving the 
equation (\ref{eqbess}) which can be rewritten as 
\begin{equation}
e^{2 \sigma}{1 \over v(\Omega)}{d \over d \Omega}\Big( v(\Omega){d \chi_n \over d \Omega} \Big) = - m_n^2\chi_n \, ,
\end{equation}
where 
\begin{equation}
v(\Omega)  =  e^{{\sigma \over w+1} [ (D-1)w+ D]}\quad ; \quad h(\Omega) =  e^{{-2\sigma \over (w+1)} [ w+ {D-1\over D-2}]}\, . \label{h(omega)}
\end{equation}

It is convenient to solve this equation in terms of the  coordinate $ z = {1\over k}e^{-\sigma}$
\begin{equation}
z^{u}
{d\over d{ z}}[ z^{-u}{d\over d{z}}\chi_n] = -m^2_n \chi_n \, ,
\end{equation}
\noindent where $u={ (D-2)\omega + D-1 \over \omega +1}$.  This equation has a zero mode solution corresponding to $m_n=0$ of the form 
\begin{equation}
\chi_n(z) = c_1 + c_2z^{u+1} \, .
\end{equation}

For $m_n>0$ the solution is a combination of BesselJ and BesselY  functions of argument $m_n z$. In terms of $\Omega$ the solution reads 
\beq
\chi_n = {e^{-\nu \sigma}\over N_n}[J_{\nu}({m_n\over k}e^{-\sigma})
+ b_{n\nu}Y_\nu ({m_n\over k}e^{-\sigma})]\, ,
\eeq
where $\sigma = - k |\Omega|$ and $\nu = {(D-1)w+D \over 2(w+1)}=(u+1)/2$ and $N_n$ is a normalization constant. Besides the condition $w >  - {D-1 \over  D-2}$, it is interesting to note that in order to find finite  $\nu$ we need $\omega \ne -1$. The limit $\omega \to \infty$  leads to the result found in \cite{Goldberger:1999wh} for the massless case.
Our modes solutions are even functions in $\Omega$. To guarantee the continuity at the orbifold points $\Omega=0$ and $\Omega=L$ we impose Neumann boundary conditions.  The boundary condition at $\Omega=0$ leads to 

\begin{figure}
\centering
\includegraphics[width=7cm]{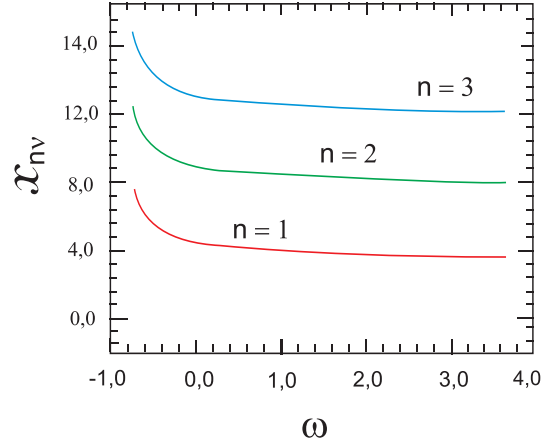}
\parbox{5in}{\caption{The curves show the  behavior of the first  Kaluza-Klein modes $x_{1 \nu}$,$x_{2 \nu}$ and $x_{3 \nu}$ as functions of the Brans-Dicke parameter  $\omega$ for the case $D=5$.
 }}
\end{figure}

 \begin{figure}
\centering
\includegraphics[width=7cm]{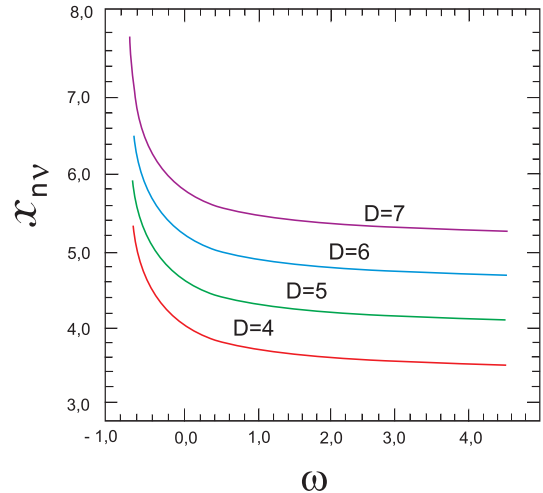}
\parbox{5in}{\caption{This figure shows the influence of the dimension $D$ on the first Kaluza-Klein mode. Each line correspond to  $x_{1 \nu}$ as a function of the Brans-Dicke parameter $\omega$  for a  particular dimension. This mode has the asymptotic values $3.14 \, , 3.83 \, , 4.49$ \, and $5.16$ for the cases $D=4,5,6$ and $D=7$ respectively.
 }\label{Figura3}}
\end{figure}
\begin{equation}
b_{n\nu} = -{ J_{\nu-1}(x_{n\nu}e^{-k L})  \over Y_{\nu-1}(x_{n\nu}e^{-k L})  } \, ,
\end{equation}
where we have defined $x_{n\nu}=(m_n/k)e^{k L}$. The boundary condition at $\Omega = \pi$ gives the important equation 
\begin{equation}
x_{n \nu}^2e^{-kL} [J_{\nu -1}(x_{n \nu})Y_{\nu -1}(x_{n \nu}e^{-kL}) - Y_{\nu -1}(x_{n \nu})J_{\nu -1}(x_{n \nu}e^{-kL})]=0\, . \label{zeroseq}
\end{equation}
The Kaluza-Klein modes  $x_{n\nu}$ are obtained by solving this equation. We choose $kL=12$ as considered in the original Randall-Sundrum model.  We present in Fig. 2 our results for the first modes as functions of the Brans-Dicke parameter $\omega$ in the particular case $D=5$ . We see from that figure that the modes grow rapidly when $\omega \to - 1$  and approach  constant functions for large $\omega$ (for instance $x_{1 \nu} \to  3.83$ for large $\omega$). This way the distance between these modes is preserved at large $\omega$.  The Fig. 3 shows how the first mode $x_{1\nu}$ increases with the dimension.

The normalization constant $N_n$ appearing in  the modes solutions can be calculated by performing the integral of eq. (\ref{normalizationcond}). This integral is not simple in general because involves products of BesselJ and BesselY functions. However, for the lower modes the dominant contribution to the integral comes from the square of BesselJ. For these cases the normalization constant can be approximated by  
\begin{equation}
N_n \approx {1 \over \sqrt{k L }}e^{kL}J_{\nu}(x_{n\nu})\, ,
\end{equation}

\noindent where we have supposed in this approximation that $kL$ is large as expected for the resolution of the hierarchy problem \cite{Mikhailov:2006vx}.

\section{The effect of the Brans-Dicke parameter in Electroweak phenomenology}

We analyze in this section an interesting application of the Brans-Dicke Randall-Sundrum scenario considered in this paper. This application concerns the Higgless model of ref. \cite{Csaki:2003zu} (a review can be found in \cite{Csaki:2005vy}). This model consists on a $SU(2)_L \times SU(2)_R \times U(1)_{B-L}$ gauge group living in a 5d AdS metric limited by flat 3-branes (the Randall-Sundrum scenario revised in section II). The gauge symmetry is broken by imposing gauge field boundary conditions on the 3-branes while the Kaluza-Klein towers arising from gauge field fluctuations are interpreted as $W^{\pm}$ and $Z$ resonances being the lowest modes associated to the experimentally observed $W$ and $Z$ particles.

In our case the metric contains an extra degree of freedom which is the Brans-Dicke parameter. As we saw in the last section, this parameter acts as a fine-tuning for the Kaluza-Klein masses arising from scalar fluctuations. We will see in this section how the $W^{\pm}$ and $Z$ resonances  of the Higgless model will depend on the Brans-Dicke parameter as well.

We begin with the $SU(2)_L \times SU(2)_R \times U(1)_{B-L}$ action 
\begin{equation}
S = -{1 \over 4} \sum_{a=1}^3 \int  d^4 x \,\int d z \sqrt{-g} \Big[  F^{M N}_{a\,\,(L) }F_{M N}^{a\,\,(L) } + F^{M N}_{a\,\,(R) } F_{M N}^{a\,\,(R) } 
+ B^{M N} B_{M N} \Big] \, , \label{higglessaction}
\end{equation}
where 
\begin{equation}
B_{M N} = \partial_{M}B_{N} - \partial_{N}B_{M} \quad ; \quad F^{a\,(L,R)}_{M N} = \partial_{M}A^a_{N} - \partial_{N}A^a_{M} + g_5 f^{a b c}A_{M }^b A_{N}^c \,,
\end{equation}
with $M=\{z, \mu \}$. We denote as $g_5$ the coupling constant of $SU(2)_L$ , $SU(2)_R$ and $\tilde g_5$ the $U(1)$ coupling constant. In order to cancel the interaction terms between the $z$ and $\mu$ components we must add gauge fixing terms of the form 
\begin{equation}
S_{gf} = {1 \over 2 \xi} \int d^4 x \, dz  \sqrt{-g}h^2(z)  \Big [ \eta^{\mu \nu} \partial_\mu {\cal A}_\nu - \frac{\xi}{\sqrt{-g} h^2(z)} \partial_z (\sqrt{-g} h^2(z) {\cal A}_z ) \Big ]^2 \,
\end{equation}
with ${\cal A}_M = \{ A_M^{a (L)},A_M^{a (R)}, B_M \}$  and 
\begin{equation}
\sqrt{-g} h^2(z) = (kz)^{-1} \, \sqrt{f_{\omega}(z)} \, . 
\end{equation}
The bulk fields can be decomposed in the following way 
\begin{eqnarray}
B_{\mu} &=& g_5 a_0 \gamma_{\mu}(x) + \sum_{n=1}^{\infty} Z_{\mu}^{(n)}(x) \psi_n^{B}(z) \, ,  \label{decompositionB} \\
A^{3 (L,R)}_{\mu} &=& \tilde g_5 a_0 \gamma_{\mu}(x) + \sum_{n=1}^{\infty} Z_{\mu}^{(n)}(x) \psi_n^{3  (L,R)}(z)  \, ,  \label{decompositionA3} \\
A^{\pm (L,R)}_{\mu} &=& \sum_{n=1}^{\infty} W_{\mu}^{\pm (n)}(x) \psi_n^{\pm (L,R)}(z) \,. \label{decompositionA12}
\end{eqnarray}
The boundary conditions on the Planck brane $z = {1 \over k}$ are 
\begin{eqnarray}
\tilde g_5 B_{\mu} - g_5 A_{\mu}^{3(R)} =0 ,   \label{conditionplanckZ2} \\
\partial_z \left[ g_5 B_{\mu} + \tilde g_5 A_{\mu}^{3(R)} \right] =0 \, , \quad   \partial_z A_{\mu}^{3L}=0 ,\label{conditionplanckZ1} \\
\partial_z A_{\mu}^{ \pm (L)}= 0 \,, \quad   A_{\mu}^{ \pm (R)} =0  \, . \label{conditionplanckW}
 \end{eqnarray}
These conditions lead to the symmetry breaking  $SU(2)_{R} \times U(1)_{B-L} \to U(1)_{Y}$.
The boundary condition at the TeV brane  $z = {1 \over k} e^{k L}$  are 
\begin{eqnarray}
A_{\mu}^{3(L)} -  A_{\mu}^{3(R)} =0 \, ,   \label{conditionTeVZ2} 
\\
\partial_z \left[ A_{\mu}^{3 (L)} + A_{\mu}^{3(R)} \right ] =  0  \, ,  \, \,\, \partial_z B_{\mu} = 0 \, ,\label{conditionTeVZ1} \\
\partial_z \left[ A_{\mu}^{ \pm (L)} + A_{\mu}^{\pm (R)} \right] =  0 \, , \, \,   \,  A_{\mu}^{\pm (L)} -  A_{\mu}^{\pm (R)} =0 \, ,
 \label{conditionTeVW}
\end{eqnarray}
that lead to  the symmetry breaking $SU(2)_L \times SU(2)_R \to SU(2)_D$. 
According to the Kaluza Klein decomposition (\ref{decompositionB}), (\ref{decompositionA3}) and (\ref{decompositionA12}) the kinetic terms read 
\begin{eqnarray}
&\,& S_{kin} = -{1 \over 4} \int  d^4 x \,dz \,\sqrt{-g} \,h^2(z)\, \Big\{ \,\eta^{\mu \alpha} \, \eta^{\nu \beta} \times \left [ (g_5^2  +  2 \tilde g_5^2) a_0 \gamma_{\mu \nu} \gamma_{\alpha \beta} +\! \! \! \!\sum_{n, m=1}^{\infty} \! \! Z_{\mu \nu}^{(n)}Z_{\alpha  \beta}^{(m)}  (\Psi^Z_n)^T \Psi^Z_m \right ] \cr 
&\! \! \! \! \! \!\! \! \!\! \! \!\! \! \!\! \!+\,& \! \! \!\! \! \! \! \! \! 2\! \! \sum_{n, m =1}^{\infty}  \eta^{\mu \nu} Z_{\mu}^{(n)} Z_{\nu}^{(m)} (\Psi_n^Z)^T \frac{ \partial_{z } \left[ \sqrt{-g}h^2(z) \partial_{z} \Psi_m^Z  \right]}{\sqrt{-g} h^2(z)} +  \sum_{a=\pm} \,\sum_{n, m=1}^{\infty}  (\Psi^{a \, W}_n)^{T} \Big[ \eta^{\mu \alpha} \eta^{\nu \beta}  W^{a\, (n)}_{\mu \nu}W^{a \, (m)}_{\alpha  \beta}   \Psi^{a \, W}_m  \nonumber\\
&\! \! \! \! \! \!\! \! \!\! \! \!\! \! \!\! \!+&\! \! \! \! \! \! \! \! \! 2 \, \eta^{\mu \nu}  W^{a\, (n)}_{\mu} W^{a \, (m)}_{\nu}\frac{\partial_{z} \left[ \sqrt{-g}h^2(z) \partial_z \Psi_m^{a \, W} \right ] }{\sqrt{-g} h^2(z)} \Big ]\Big\} 
\end{eqnarray}
where we defined the vectors $\Psi^Z_n \equiv \{ \psi^{(B)}_n , \psi^{3(L)}_n ,
\psi^{3 (R)}_n \} $ and $\Psi^{W}_n \equiv \{ \psi^{a (L)}_n ,  \psi^{ a(R)}_n \} $.  
This decomposition suggests the normalization conditions 
\begin{eqnarray}
a_0 (g_5^2 + 2 \tilde g_5^2)\int_{-L}^{L} dz \sqrt{-g} h^2(z) = 1 \, , \cr
\int^L_{-L} d z \sqrt{-g} h^2(z) \Psi_n^{ Z \, T } \Psi^Z_m = \delta_{m n} \, , 
\int^L_{-L} d z \sqrt{-g} h^2(z) (\Psi_n^{a \, W})^{ T } \Psi^{a \, W}_m = \delta_{m n} \, ,
 \label{normalizationHigglessmodel}
\end{eqnarray}
and the following equation of motion 
\begin{eqnarray}
z^{\bar u}\frac{d}{d z}\Big[ z^{-\bar u}\frac{d}{d z} \Psi^Z_n \Big] &\,=\,& - (m_n^{Z})^{2} \, 
 \Psi_n^Z \, , \cr 
z^{\bar u}\frac{d}{d z}\Big[ z^{-\bar u}\frac{d}{d z} \Psi^{ a \, W} _n \Big] &\,=\,& - (m_n^{a \, W})^{2} \, 
 \Psi_n^{a \, W} \, , \label{equationmodes}
\end{eqnarray}
where $\bar u = \frac{\omega + 4/3 }{\omega + 1 }$. 
The solution to equation (\ref{equationmodes}) is
\begin{equation}
\psi_n (z)  = {(kz)^{\bar \nu} \over N_n} \Big[J_{\bar \nu}( m_n z) + b_{n \bar \nu} Y_{\bar \nu} (m_n z) \Big] \, ,
\end{equation}
where $\bar \nu = \frac{\omega + 7/6 }{\omega + 1 }$ and $i=\{Z , \, a \, W \}$. By substituting the decompositions (\ref{decompositionB}) and (\ref{decompositionA3}) into the boundary conditions (\ref{conditionplanckZ1}),(\ref{conditionplanckZ2}), (\ref{conditionTeVZ1}) and (\ref{conditionTeVZ2}) we obtain the mass equation for the boson Z :
\beq
(R_{ \bar \nu \!   -  \! 1}\! - \!\tilde R_{\bar \nu  - \! 1})(R_{\bar \nu} -\! \tilde R_{\bar \nu}) + (R_{\bar \nu  - \!1}  \!-\! \tilde R_{\bar \nu} )(R_{\bar \nu} \!- \! \tilde R_{\bar \nu \! - \!1})    +  2 \, \, \frac{\tilde g_5^2}{g_5^2 } (R_{\bar \nu - 1} -  \bar R_{\bar \nu})(R_{\bar \nu} -\tilde R_{\bar \nu -1}) = 0 \, ,
\label{RZ}
\eeq
where 
\begin{equation}
R_{\bar \alpha}  =  -{J_{\bar \alpha}(x_{n {\bar \nu}}\, e^{-k L}) \over Y_{\bar \alpha}(x_{n {\bar \nu}}\, e^{-k L})} \quad , \quad \tilde R_{\bar \alpha}  =  -{J_{\bar \alpha}(x_{n {\bar \nu}}) \over Y_{{\bar \alpha}}(x_{n {\bar \nu}})} \, ,
\end{equation}
with $x_{n {\bar \nu}}=(m_n/k)\,e^{kL}$, $\bar \alpha= \{ \bar \nu, \bar \nu -1 \}$ and we assumed that $g_5^2>0$. Similarly, substituting  (\ref{decompositionA12}) into (\ref{conditionplanckW}) and (\ref{conditionTeVW}) , we find the $W^{\pm}$  mass equation
\begin{eqnarray}
(R_{ \bar \nu \!  -  \! 1}\! - \! \tilde R_{\bar \nu \! - \! 1})(R_{\bar \nu} \! \!-\! \tilde R_{\bar \nu}) \!+ \! (R_{\bar \nu \! - \!1}  \!-\! \tilde R_{\bar \nu} )(R_{\bar \nu} \!- \! \tilde R_{\bar \nu \! - \!1} ) = 0 \, .\label{RW}
\end{eqnarray}

\begin{figure}
\centering
\includegraphics[width=7cm]{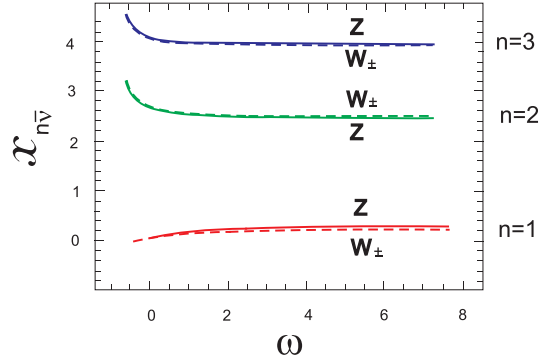}
{\caption{Kaluza-Klein modes for the Z  and $W_{\pm}$ bosons as a function of the  Brans-Dicke parameter $\omega$  for $kL=12$ and  $\tilde g_5^2/ g_5^2=0.426$. A similar behavior is obtained for other values of $\tilde g_5^2/ g_5^2$.
 }\label{FiguremassZandW}}
\end{figure}

The mass equations (\ref{RZ}) and (\ref{RW}) reduce to the usual Higgless model \cite{Csaki:2003zu} for $\bar \nu = 1$. The main difference here is that the index $\bar \nu$ varies with the  Brans-Dicke parameter $\omega$ so that the $Z$ and $W$ boson masses depend on $\omega$ as well. By numerical analysis of eq. (\ref{RW}) we conclude that the effect of the Brans-Dicke parameter $\omega$ is the following : {\it when decreasing $\omega$ the masses of the first W and Z modes decrease while  the mass of the higher modes increase}. This behavior is shown in Figure \ref{FiguremassZandW} for $kL=12$ and  $\tilde g_5^2/ g_5^2=0.426$. Note that when $\omega \to -1$ the first mode vanishes while the higher modes diverge.

\begin{figure}
\centering
\includegraphics[width=7cm]{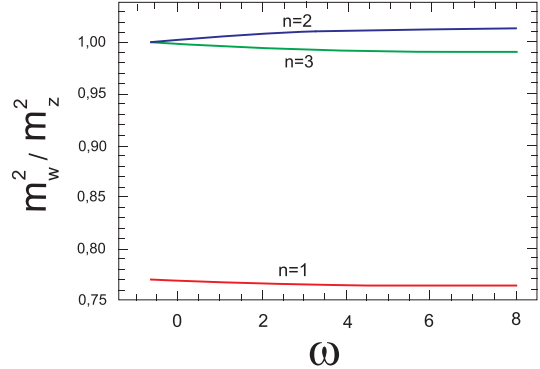}
{\caption{Quotient $m^2_W/m^2_Z$ as a function of the Brans-Dicke parameter $\omega$ for $kL=12$ and  $\tilde g_5^2/ g_5^2=0.426$. A similar behavior is obtained for other values of $\tilde g_5^2/ g_5^2$.}
\label{quocientmassWmassZ}}
\end{figure}

Another interesting result is the evolution of the quotient  $m^2_W/m^2_Z$ with the Brans-Dicke parameter where $m_W$ and $m_Z$ are the masses of the W and Z resonances.  This quotient is lower than 1 for the first and third modes and increases when decreasing $\omega$ while for the second mode it is greater than 1 and decreases when decreasing $\omega$. This behavior is shown in figure \ref{quocientmassWmassZ} for $kL=12$ and  $\tilde g_5^2/ g_5^2=0.426$. 

These values were chosen to obtain a realistic value for  the  quotient $m^2_W/m^2_Z$ in the limit $\omega \to \infty$. Indeed, in this limit we obtain  the result $m^2_W/m^2_Z \sim 0.764$ that can be compared with the  asymptotic expression 
\beq
\frac{m^2_W}{m^2_Z} \approx \frac{1 + \frac{\tilde g_5^2}{g_5^2}}{1 + 2\frac{\tilde g_5^2}{g_5^2}} \sim 0.770 \, , 
\eeq
obtained in \cite{Csaki:2003zu} for the limits $kL \gg 1$ and $k \gg 1$. According to \cite{Csaki:2003zu} (and also \cite{Csaki:2005vy}) we can also relate the couplings $g_5$ and $\tilde g_5$ to the effective Standard Model couplings $g$ and $g'$ by adding matter fields. This leads to the asymptotic relations 
\beq
g^2 \approx \frac{g_5^2}{L}  \quad , \quad g'^2 \approx \frac{g_5^2 \tilde g_5^2}{L (g_5^2 + \tilde g_5^2)}  \, . 
\eeq
Then for $\tilde g_5^2/ g_5^2=0.426$ we obtain 
\beqa
\tan^2 \theta_W &=& \frac{g'^2}{g^2}=\frac{\frac{\tilde g_5^2}{g_5^2}}{1 + \frac{\tilde g_5^2}{g_5^2}} \approx 0.299 \, , \\
\cos^2 \theta_W  &\approx& \frac{1 + \frac{\tilde g_5^2}{g_5^2}}{1 + 2 \frac{\tilde g_5^2}{g_5^2}} \approx \frac{m^2_W}{m^2_Z} \, , \label{rhoparam}
\eeqa
where $\theta_W$ is  the Weinberg angle. The relation (\ref{rhoparam}) is characteristic of Higgless models that preserve the $SU(2)$ custodial symmetry. 

\section{Conclusions}

In this paper we have constructed D-dimensional Randall-Sundrum models from Brans-Dicke theory by using a BPS-like mechanism for solving the background equations. We have also studied the Kaluza-Klein decomposition of massless scalar and gauge fields  and showed how the Kaluza-Klein modes depend on the Brans-Dicke parameter $\omega$. In particular, we saw how the Brans-Dicke parameter act as a fine-tuning parameter for the W and Z resonances of a Higgless electroweak model. 

We have considered in our analysis of scalar and vector Kaluza-Klein modes a wide range of values for the Brans-Dicke parameter $\omega$. We also assumed that $kL$ is large as is expected for solving the Planck-weak hierarchy problem. However, it is important to remark that stability of this model requires the addition of scalar field potentials on the Planck and TeV branes.  As mentioned in \cite{Mikhailov:2006vx}, after introducing stabilizing potentials a large value of $\omega$ is needed
in order to avoid a new hierarchy for the scalar field.

In the Brans-Dicke theory the presence of a background scalar field was crucial. A possible future investigation would be studying the effect of other background fields like the Kalb-Ramond field which is motivated by String Theory (see for instance \cite{Mukhopadhyaya:2004cc,Braga:2004ns}). 

Another interesting feature to be explored is the effect of the Brans-Dicke parameter on scalar and gauge field interactions and in the presence of fermionic fields. .

\vspace{.3 true cm}

\noindent {\bf Acknowledgments}: The authors are financially supported by CNPq.

\end{document}